\documentclass[sigconf,natbib=true,anonymous=false]{acmart}
\AtBeginDocument{%
  \providecommand\BibTeX{{%
    \normalfont B\kern-0.5em{\scshape i\kern-0.25em b}\kern-0.8em\TeX}}}

\setcopyright{acmcopyright}

\copyrightyear{2022}
\acmYear{2022}
\setcopyright{acmcopyright}\acmConference[SIGIR '22]{Proceedings of the 45th International ACM SIGIR Conference on Research and Development in Information Retrieval}{July 11--15, 2022}{Madrid, Spain}
\acmBooktitle{Proceedings of the 45th International ACM SIGIR Conference on Research and Development in Information Retrieval (SIGIR '22), July 11--15, 2022, Madrid, Spain}
\acmPrice{15.00}
\acmDOI{10.1145/3477495.3531902}
\acmISBN{978-1-4503-8732-3/22/07}
\usepackage[linesnumbered,lined]{algorithm2e}
\usepackage{booktabs}
\usepackage{bookmark,xspace}
\usepackage{multirow}
\usepackage{paralist}
\usepackage{graphicx}
\usepackage{subfigure}
\usepackage{amsmath, nccmath}
\usepackage{color}
\usepackage{xcolor}
\usepackage{arydshln}

\begin{document}

\title{Improving Item Cold-start Recommendation via Model-agnostic Conditional Variational Autoencoder}

\author{Xu Zhao}
\affiliation{%
  \institution{Tencent News}
  \city{Beijing}
  \country{China}}
\email{xuzzzhao@tencent.com}
\orcid{0000-0001-5146-5789}

\author{Yi Ren}
\affiliation{%
  \institution{Tencent News}
  \city{Beijing}
  \country{China}}
\email{henrybjren@tencent.com}

\author{Ying Du}
\affiliation{%
  \institution{Tencent News}
  \city{Beijing}
  \country{China}}
\email{yingdu@tencent.com}

\author{Shenzheng Zhang}
\affiliation{%
  \institution{Tencent News}
  \city{Beijing}
  \country{China}}
\email{qjzcyzhang@tencent.com}

\author{Nian Wang}
\affiliation{%
  \institution{Tencent News}
  \city{Beijing}
  \country{China}}
\email{noreenwang@tencent.com}

%

\renewcommand{\shortauthors}{Xu Zhao, et al.}





\begin{abstract}
    Embedding \& MLP has become a paradigm for modern large-scale recommendation system. However, this paradigm suffers from the cold-start problem which will seriously compromise the ecological health of recommendation systems. This paper attempts to  tackle the item cold-start problem by generating enhanced warmed-up ID embeddings for cold items with historical data and limited interaction records. From the aspect of industrial practice, we mainly focus on the following three points of item cold-start: 1) How to conduct cold-start without additional data requirements and make strategy easy to be deployed in online recommendation scenarios. 2) How to leverage both historical records and constantly emerging interaction data of new items. 3) How to model the relationship between item ID and side information stably from interaction data. To address these problems,  we propose a model-agnostic \textbf{C}onditional \textbf{V}ariational \textbf{A}utoencoder based  \textbf{R}ecommendation(\name) framework with some advantages including compatibility on various backbones, no extra requirements for data, utilization of both historical data and recent emerging interactions. \name uses latent variables to learn a distribution over item side information and generates desirable item ID embeddings using a conditional decoder. The proposed method is evaluated by extensive offline experiments on public datasets and online A/B tests on Tencent News recommendation platform, which further illustrate the advantages and robustness of \name.

\end{abstract}




\newcommand{\name}{CVAR\xspace}
\newlength\savedwidth
\newcommand\whline{\noalign{\global\savedwidth\arrayrulewidth
                           \global\arrayrulewidth 1.5pt}%
                  \hline
                  \noalign{\global\arrayrulewidth\savedwidth}}

\fancyhead{}

\maketitle

\section{Introduction}

\begin{figure}

    \setlength{\belowcaptionskip}{-0.4 cm}
    \centering
\scalebox{0.9}{
    \includegraphics[width=\linewidth]{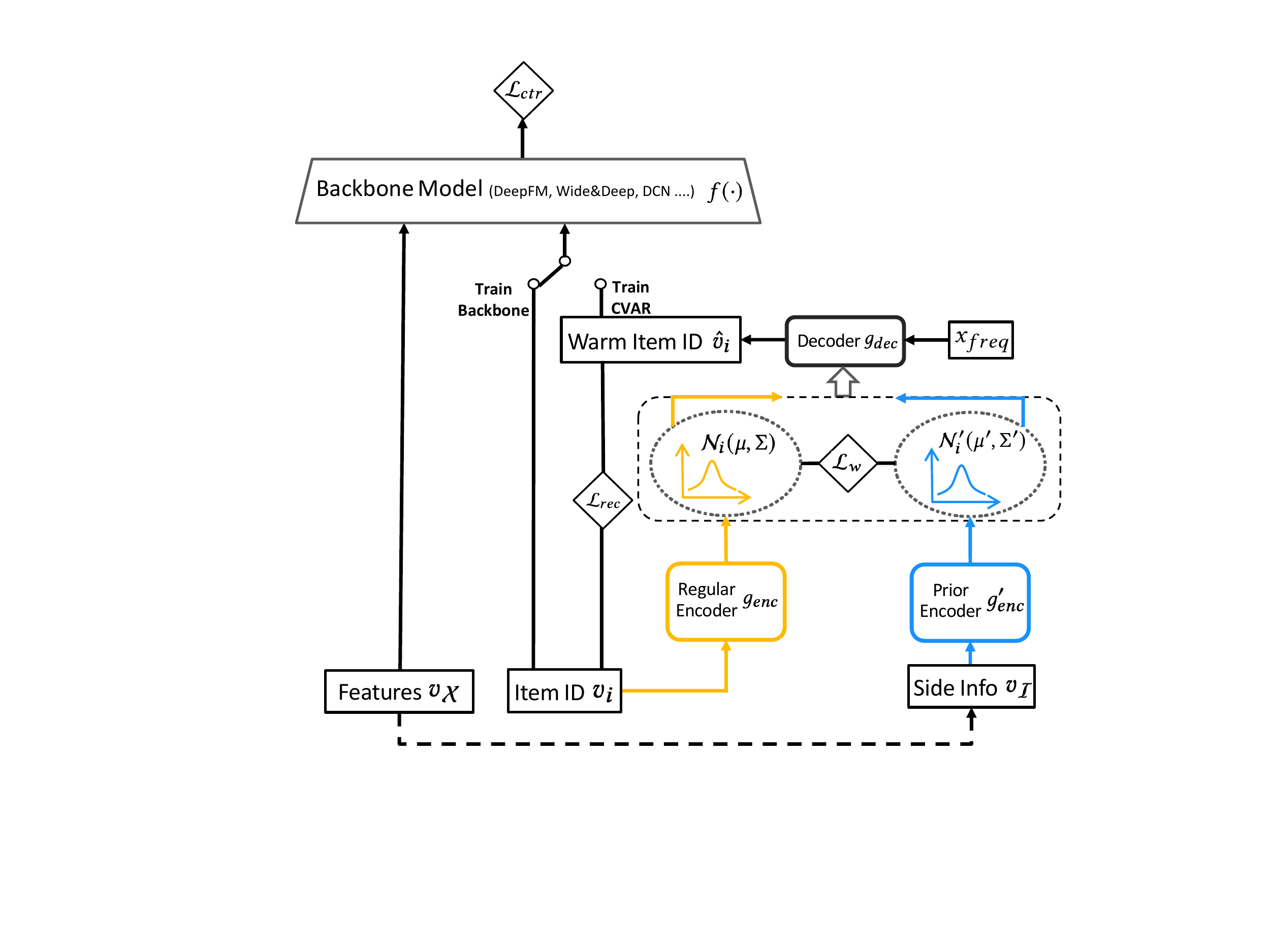}}
    \caption{The proposed model-agnostic \name Framework, where Backbone Model can be any CTR prediction model with embedding layer.}
    \label{fig:framework}
\vspace{-0.2cm}
\end{figure}

In the era of mobile internet, various online applications continuously emerge and explosively grow, in which recommendation systems play a key role in connecting users and content. 
Due to the excellent scalability and convenience of handling massive features, Embedding \& MLP~\cite{sedhain2015autorec,he2017neural,guo2017deepfm,cheng2016wide} has become a paradigm for modern large-scale recommendation systems~\cite{zhou2018deep,ma2018modeling,covington2016deep,zhou2019deep,davidson2010youtube}.
However, this paradigm is data demanding and suffers from cold-start problem~\cite{lam2008addressing,sethi2021cold,gope2017survey}. Concretely, for a large amount of new emerging items with limited interactions, their embeddings are insufficiently trained which leads to poor recommendation performance.
The cold-start problem has become a crucial obstacle for online recommendation. Under the influence of Pareto~\cite{ribeiro2014multiobjective} effect in our industrial system, A small portion of well-trained items tend to obtain more accurate recommendations and more impressions,  which will further compromise the distributing efficiency of system. 

Some approaches have been proposed to address the item cold-start challenge. CLCRec~\cite{wei2021contrastive} proposes to address cold-start problem by maximizing two kinds of mutual information using contrastive learning technology. Heater~\cite{zhu2020recommendation} uses the sum squared error (SSE) loss to model the collaborative embedding. Notice that CLCRec and Heater are only designed for the CF-based backbone models. There are also some model-agnostic methods that could be widely equipped to various backbones. 
DropoutNet~\cite{volkovs2017dropoutnet} applies dropout technology to relieve the model‘s dependency on item ID. Meta Embedding~\cite{pan2019warm} focuses on learning how to learn the ID embedding for new items with meta learning technology. MWUF~\cite{zhu2021learning} proposes meta Scaling and meta Shifting Networks to warm up cold ID embeddings. Although these methods are model-agnostic, they all have extra and strict requirements on data. For instance, DropoutNet and MWUF require the interacted user set of items for cold-start, while Meta Embedding requires building two mini-batches containing the same items for training. In the online scenario of industrial recommendation, these extra requirements on data stream make the deployment process rather difficult.

Generally, there are two ways to solve the cold-start problem: the first is to mine the distribution patterns hidden in historical data~\cite{zheng2021multi, wang2020disenhan,pan2019warm,zhu2021learning,zhao2020catn}, such as learning the transformation relationship between side information and item ID~\cite{wei2021contrastive,zhu2020recommendation,pan2019warm,zhu2021learning}. The second is to improve the learning efficiency with limited samples of cold items, such as methods based on meta learning~\cite{lee2019melu, zhang2021model, ouyang2021learning,lu2020meta}. However, previous works rarely consider both directions at the same time. In other words, methods in the first category only focus on the initialization of embedding, while the methods in the second category usually ignore patterns hidden in historical data.


Considering the issues mentioned above in previous research along with our industrial practice, we summarize three key points for the design of cold-start method: 
1) How to conduct cold-start without additional data requirements and make strategy easy to be deployed in online recommendation scenario.
2) How the cold-start method can leverage both historical records and constantly emerging interaction data of new items.
3) How to model the relationship between item ID and side information stably from interaction data and minimize the discrepancy between cold item ID embedding and fully trained embedding space.

To achieve these desiderata, we propose a \textbf{C}onditional \textbf{V}ariational \textbf{A}utoencoder based model-agnostic \textbf{R}ecommendation (\name) framework. As an independent framework, \name could be equipped on various backbone models and be trained in an end-to-end way, using the same samples as backbone. Thus \name makes no redundant requirements for training data. In addition to giving desirable initialization in the cold-start phase, \name will leverage the continuously updating item ID embeddings from the backbone to generate enhanced warmed-up embeddings with superior quality. Therefore \name can fulfill the second concern mentioned above. 

As for the third problem mentioned before, \name aligns the representation of item ID and side information 
in the latent space of \textbf{C}onditional \textbf{V}ariational \textbf{A}uto\textbf{E}ncoder(CVAE)~\cite{DBLP:conf/nips/SohnLY15,DBLP:conf/acl/ZhaoZE17,walker2016uncertain}, which is shown in Figure~\ref{fig:framework}.
Specifically, previous works~\cite{zhu2020recommendation,pan2019warm,zhu2021learning} usually train a learnable mapping from item side information to item ID. 
However, item ID contains not only content information, but also lots of interaction information which makes it difficult to learn a precise mapping directly. Inspired by Denoise Autoencoder~\cite{vincent2010stacked}, we first conduct dimension reduction of item ID embedding with Encoder-Decoder paradigm and get the denoised representation of item ID in a latent space. Then the side information is transformed to the latent space and aligned with the denoised representation of item ID. Corresponding with the design in CVAE, latent representation is defined as normal distribution which could maintain some Exploit-Exposure ability for \name.


The main contributions of this work are summarized into four folds:
\vspace{-0.1cm}
\begin{enumerate}
    \item We propose a model-agnostic \name to warm up cold item ID embeddings. \name has no extra data requirements which makes it easy to be deployed in online scenario.
    \item \name not only learns the pattern in historical data but also leverages the continuously updating item ID embedding from the backbone to generate enhanced warmed-up embeddings with superior quality.
    \item We propose to model the relationship between item id and side information in the latent space and generate desirable ID embeddings using a conditional decoder.
    \item Extensive offline and online experiments are conducted to demonstrate the effectiveness and compatibility of \name.
\end{enumerate}

\vspace{-0.2cm}
\section{Proposed Method}

In this section, we propose \name, a model-agnostic framework to warm up ID embeddings for new items.  
\name is designed based on the \textbf{C}lick-\textbf{T}hrough-\textbf{R}ate(CTR) prediction task~\cite{richardson2007predicting, graepel2010web}, predicting the click/watch/purchase behavior in recommendation scenario, which is usually formulated as a supervised binary classification task. Each sample in CTR task consists of multiple input features $\mathbf{x}$ and the binary label $y$. Generally, the input features x could be splitted into three parts, i.e. $\mathbf{x}=(i, \mathcal{X})$. 
\begin{itemize}
    \item $i$, item ID, a unique number or string to identify each item in recommendation system.
    \item $\mathcal{X}=\{x_1, ..., x_{|\mathcal{X}|}\}$, features which are used for CTR predict , may be categorical or continuous, such as the user attributes and contextual information.
\end{itemize}

Moreover, We take part of features from $\mathcal{X}$ as item side information $\mathcal{I} \subset \mathcal{X}$, which will be consumed in item cold-start procedure. 
Standard feature preprocessing~\cite{richardson2007predicting, graepel2010web} has been applied in this work. Continuous features are normalized to range between 0 and 1. Following the embedding technology~\cite{DBLP:conf/icml/LeM14, DBLP:journals/corr/abs-1301-3781}, categorical features are transformed to dense vectors, called embeddings. Normalized values of continuous features and dense embeddings of categorical features are concatenated together to constitute the final representation of input features. We denote the representations of $i$, $\mathcal{I}$, $\mathcal{X}$ as $v_{i} \in \mathbb{R}^d$, $v_{\mathcal{I}} \in \mathbb{R}^{d \times |\mathcal{I}|}$ and $v_{\mathcal{X}} \in \mathbb{R}^{d \times |\mathcal{X}|}$ respectively. 
Note that $v_{\mathcal{I}}$ and $v_{\mathcal{X}}$ are the concatenation of multiple feature embeddings. 

The CTR target is to approximate the probability $\hat{y} = Pr(y=1|\mathbf{x})$ by a discriminative function $f(\cdot)$:
\begin{equation}
\setlength\abovedisplayskip{3pt}
\setlength\belowdisplayskip{3pt}
\hat{y} = f(v_i, v_{\mathcal{X}};\theta) \label{equ:backbone}
\end{equation}
where $\theta$ denotes the parameters of the backbone model $f(\cdot)$. Then the Binary Cross Entropy~\cite{de2005tutorial} is used to format the loss function:
\begin{equation}
\setlength\abovedisplayskip{3pt}
\setlength\belowdisplayskip{3pt}
    \mathcal{L}(\theta, \phi)=-y \log \hat{y}-(1-y) \log (1-\hat{y})
\end{equation}
where $\phi$ denotes the parameters of the embedding layer, including $v_i$, $v_{\mathcal{I}}$ and $v_{\mathcal{X}}$. 
As the parameters is trained by historical data, the item ID embeddings of recent emerging items are rarely updated and stay around the initial point, leading to a low testing accuracy. It is known as the item cold-start problem.

As the relationship between cold-start and warm-up in recommendation system is confusing, we will give a brief discussion here. Generally speaking, cold-start strategies are applied to items which first appear in the system, while warm-up strategies are applied to items whose exposure numbers are lower than a threshold. Note that this threshold is inconsistent across different systems. In other words, warm-up is a subsequent procedure of cold-start. In this work, We conduct cold-start and warm-up in a unified framework \name. 

The structure of \name is shown in Figure~\ref{fig:framework}.
The basic design of \name is generating a better embedding for item ID and replace the original unsufficiently trained embedding. 
The Item ID in Figure~\ref{fig:framework} denotes the original item ID embedding $v_i$ trained by backbone model, while the Warm Item ID denotes the enhanced warmed-up embedding $\hat{v}_i$ generated by \name. 
Instead of directly learning a transformation from $v_{\mathcal{I}}$ to $v_i$, \name aligns the representations transformed from $v_i$ and $v_{\mathcal{I}}$ in the autoencoder's latent space~\cite{wang2016auto, durkan2019neural,ye2021autoencoder}. Following the design of CVAE~\cite{DBLP:conf/nips/SohnLY15}, the autoencoder applied to $v_i$ is formulated as:
\begin{align}
\setlength\abovedisplayskip{1pt}
\setlength\belowdisplayskip{1pt}
    &\mu, \sigma = g_{enc}(v_i; w_{enc}); \mu \in \mathbb{R}^k, \sigma \in \mathbb{R}^k \label{equ:reg_enc}\\
    &z \sim \mathcal{N}_i(\mu, \Sigma); \Sigma \in \mathbb{R}^{k \times k}, diag(\Sigma) = \sigma \label{equ:reparam}\\
    &\hat{v}_{i} = g_{dec}(z, x_{freq}; w_{dec}); \hat{v}_{i} \in \mathbb{R}^d \label{equ:warm}
\end{align}
where $g_{enc}$ and $g_{dec}$ correspond to the Regular Encoder and Decoder in Figure~\ref{fig:framework}, $w_{enc}$ and $ w_{dec}$ denote the parameters of $g_{enc}$ and $g_{dec}$, $k$ denotes the dimension of latent space. Notice that as shown in Equation~\eqref{equ:reg_enc} the latent representation is defined as a multivariate normal distribution $\mathcal{N}_i(\mu, \Sigma)$ with mean $\mu$ and diagonal covariance matrix $\Sigma$ whose trace is $\sigma$. In Equation~\eqref{equ:reparam}, latent representation $z$ is sampled from $\mathcal{N}_i(\mu, \Sigma)$ using the reparameterization trick~\cite{DBLP:conf/coling/PhamL20}. Decoder $g_{dec}$ takes  $z$ along with conditional frequency $x_{freq}$ as input and reconstructs item ID embedding as $\hat{v}_i$. 
Since the item frequency information has a direct impact on the distribution of ID embedding and \name is operated in the entire warm-up phase, unlike traditional CVAE design~\cite{DBLP:conf/nips/SohnLY15,DBLP:conf/acl/ZhaoZE17,walker2016uncertain}, we only filter $x_{freq}$ from the full side information as the condition of decoder to emphasize its impact. 

Reconstruction Loss between $v_i$ and $\hat{v}_i$ is formulated by Euclidean distance:
\begin{align}
\label{equ-rec}
\setlength\abovedisplayskip{1pt}
\setlength\belowdisplayskip{1pt}
    \mathcal{L}_{rec}(w_{enc}, w_{dec}) = \|v_i - \hat{v}_i\|^2_2
\end{align}
Notice that rarely updated $v_i$ of almost cold items in Equation~\eqref{equ-rec} may mislead the training procedure. However,
with $x_{freq}$ as condition to $g_{dec}$, the generated embedding will be restricted to a reasonable space according to $x_{freq}$ which could relieve this misleading effect. Thus we do not deal with this case separately. Besides, in inference stage, we set $x_{freq}$ as a huge value which will help to produce warm ID embeddings.

Through the autoencoder structure described above,  we obtain the information-compressed latent distribution $\mathcal{N}_i(\mu, \Sigma)$ of $v_i$. Aligning embeddings or distributions in autoencoder's latent space is a generally used technology which has been proven effective in various fields~\cite{durkan2019neural}. Thus we consider mapping the side information embedding $v_{\mathcal{I}}$ into the same latent space and aligning with $\mathcal{N}_i(\mu, \Sigma)$. Specifically, Prior Encoder $g^{\prime}_{enc}$ maps $v_{\mathcal{I}}$ to $\mathcal{N}_{i}(\mu^{\prime}, \Sigma^{\prime})$ and a wasserstein loss~\cite{peyre2019computational, shen2018wasserstein} function is applied to aligning distribution $\mathcal{N}_{i}(\mu, \Sigma)$ with $\mathcal{N}_{i}(\mu^{\prime}, \Sigma^{\prime})$:
\begin{align}
\setlength\abovedisplayskip{1pt}
\setlength\belowdisplayskip{1pt}
    &\mu^{\prime}, \sigma^{\prime} = g^{\prime}_{enc}(v_{\mathcal{I}}; w^{\prime}_{enc}); \mu^{\prime} \in \mathbb{R}^k, \sigma^{\prime} \in \mathbb{R}^k \label{equ:pri_enc} \\
    &z^{\prime} \sim \mathcal{N}_{i}^{\prime}(\mu^{\prime}, \Sigma^{\prime}); \Sigma^{\prime} \in \mathbb{R}^{k \times k}, trace(\Sigma^{\prime}) = \sigma^{\prime} \label{equ:reparam_1}\\
    &\hat{v}_{i}^{\prime} = g_{dec}(z^{\prime}, x_{freq}; w_{dec}); \hat{v}_{i}^{\prime} \in \mathbb{R}^d \label{equ:warm_1}\\
     &\mathcal{L}_{w}(w_{enc}, w^{\prime}_{enc}) = W_2(\mathcal{N}_{i}(\mu, \Sigma), \mathcal{N}_{i}^{\prime}(\mu^{\prime},
     \Sigma^{\prime}))
\end{align}
where $w^{\prime}_{enc}$ denotes the parameters of $g^{\prime}_{enc}$, $W_2(\cdot, \cdot)$ denotes the Wasserstein Distance. Instead of KL Divergence~\cite{bu2018estimation} or other distribution measurement, we choose Wasserstein Distance considering its symmetry and stability which is widely used in various scenarios~\cite{zhao2020relaxed,zhao2020semi,wang2020robust,liao2022fast}.

We will give an additional discussion about the design of $x_{freq}$.
Since that it's not feasible to extract frequency information from side information by $g^{\prime}_{enc}$ in item cold start scenario, the latent space ideally should contain no frequency information.
However, the frequency information is already utilized in the origin item ID embedding $v_{i}$ and automatically compressed to the latent space under the Encoder-Decoder framework. 
Thus we set $x_{freq}$ as the a independent condition of decoder to reduce the proportion of frequency information in latent space.

In addition to $\mathcal{L}_{rec}$ and $\mathcal{L}_{w}$,  we replace $v_i$ in Equation~\eqref{equ:backbone} with enhanced warmed-up $\hat{v}_{i}^{\prime}$ as item ID input to backbone and get the CTR loss $\mathcal{L}_{ctr}$ by forward computation:
\begin{align}
\setlength\abovedisplayskip{1pt}
\setlength\belowdisplayskip{1pt}
    \hat{y}_{warm} &= f(\hat{v}_i^{\prime}, v_{\mathcal{X}};\theta) \\
    \mathcal{L}_{ctr}(w_{enc}, w_{dec}) = -y \log& \hat{y}_{warm}-(1-y) \log (1-\hat{y}_{warm}) \label{equ:warm_ctr_loss}
\end{align}
To avoid disturbing the recommendation of hot items, \name is taken as an independent module with backbone. During the training of \name, optimization of $\mathcal{L}_{ctr}$ in Equation~\eqref{equ:warm_ctr_loss} is 
only applied to $w_{enc}$ and $w_{dec}$, while the parameters $\theta$ of $f(\cdot)$ are fixed. We finally get the loss function to train \name:
\begin{align}
\setlength\abovedisplayskip{1pt}
\setlength\belowdisplayskip{1pt}
    \mathcal{L}_{warm}(w_{enc}, w_{dec}, w^{\prime}_{enc}) = \mathcal{L}_{ctr} + \alpha \mathcal{L}_{rec} + \beta \mathcal{L}_{w} \label{equ:warm_loss}
\end{align}
where $\alpha$ and $\beta$ are hyperparameters to fuse the losses.
Moreover, training of \name is along with the backbone's training.
As shown in Equation~\eqref{equ:warm_loss}, \name will be trained by optimizing $w_{enc}, w_{dec}, w^{\prime}_{enc}$ to minimize $\mathcal{L}_{warm}$, using the same samples as training backbone, without any additional requirements on data. For a coming batch of samples, it's first fed to backbone to update the original item ID embedding $v_i$, then used to train \name and update $w_{enc}, w_{dec}, w^{\prime}_{enc}$. Updated $v_i$ is also consumed in the training of \name at each step. Thus we claim that \name not only learns the pattern in historical data but also uses the information of update at each step to relieve the cold-start issue.

In inference phase of recent emerging items, their item ID embeddings $v_i$ are not well tained. Therefore we obtain $z{\prime}$ by sampling from $\mathcal{N}^{\prime}_{i}$ which is generated by item side information and get enhanced warmed-up ID item $\hat{v}_{i}^{\prime}$ by passing $z^{\prime}$ to $g_{dec}$ as Equation~\eqref{equ:pri_enc} and ~\eqref{equ:reparam_1} shown. Then we could replace $v_i$ with $v_i^{\prime}$ for testing of recent emerging items. As marked in Figure~\ref{fig:framework}, Equation~\eqref{equ:reparam} and~\eqref{equ:warm} are operated in training of \name, while Equation~\eqref{equ:reparam_1} and~\eqref{equ:warm_1} play a role for recent emerging items in testing phase. For a single item, besides just searching for a better initialization of ID embedding, this replacement operation will continue until $v_i$ is fully trained.

\section{Experiments}
\begin{table}[]
    \tiny
    \setlength{\abovecaptionskip}{3pt}
    \caption{Model Comparison of cold-start effectiveness on two datasets(\textit{Movielens 1M} and \textit{Taobao Ad}), under two backbones(\textit{DeepFM} and \textit{Wide\&Deep}), three runs for each. The best improvements are highlighted in bold.}
    \label{tbl-cold-start}
    \centering
    \renewcommand{\arraystretch}{1.2}
    \begin{tabular}{c|cc|cc|cc|cc}
        \whline
        \multirow{2}{*}{Methods} & \multicolumn{2}{c|}{Cold phase} & \multicolumn{2}{c|}{Warm-a phase} & \multicolumn{2}{c|}{Warm-b phase} & \multicolumn{2}{c}{Warm-c phase} \\
         & AUC & F1 & AUC & F1 & AUC & F1 & AUC & F1           \\ \hline
        \multicolumn{9}{l}{\textit{\textbf{Dataset: Movielens 1M \& Backbone: DeepFM}}} \\
        DeepFM &0.7267&	0.6231&	0.7424&	0.6383&	0.7574&	0.6503&	0.7694&	0.6608 \\
        DropoutNet& 0.7387 & 0.6339 &	0.7491 &	0.6441 &	0.7587 	&0.6531 &	0.7673 &	0.6599 \\
        Meta-E& 0.7327 &	0.6344 &	0.7441 &	0.6432 &	0.7544 	&0.6519 &	0.7633 &	0.6592  \\
        MWUF& 0.7316 &	0.6289 &	0.7462 &	0.6413 &	0.7589 &	0.6521 &	0.7701 &	0.6616 \\ \cdashline{1-9}[1.5pt/2pt]
        \name(Init Only)& 0.7401 &	0.6353 &	0.7518 &	0.6454 	&0.7624 &	0.6547 &	0.7717 &	0.6622 \\ 
        \name & \textbf{0.7419} &	\textbf{0.6356} &	\textbf{0.7927} &	\textbf{0.6789} &	\textbf{0.8021} & 	\textbf{0.6856} &	\textbf{0.8041} 	&\textbf{0.6878}  \\ 
        \hline
        \multicolumn{9}{l}{\textit{\textbf{Dataset: Movielens 1M \& Backbone: Wide\&Deep}}} \\
        Wide\&Deep &0.7071 &	0.5972 &	0.7232 &	0.6164 	&0.7354 &	0.6273 &	0.7461 	&0.6372 \\
        DropoutNet& \textbf{0.7125} &	\textbf{0.6038} &	0.7228 &	0.6159 &	0.7313 	&0.6244 &	0.7390 	&0.6314 \\
        Meta-E& 0.6727 	&0.5287 &	0.7201 &	0.6120 &	0.7345 &	0.6280 &	0.7450 	& 0.6374   \\
        MWUF& 0.7063 &	0.5966 &	0.7230 &	0.6157&	0.7355& 	0.6275& 	0.7459& 	0.6366 \\ \cdashline{1-9}[1.5pt/2pt]
        \name(Init Only)& 0.7020 &	0.5795 &	0.7255 &	0.6160& 	0.7375& 	0.6293& 	0.7473 &	0.6380  \\ 
        \name & 0.6937 &	0.5643 &	\textbf{0.7627} &	\textbf{0.6525} &	\textbf{0.7756} &	\textbf{0.6639} &	\textbf{0.7840} &	\textbf{0.6712}  \\
        \hline
        \multicolumn{9}{l}{\textit{\textbf{Dataset: Taobao Ad \& Backbone: DeepFM}}} \\
        DeepFM & 0.5983 & 0.1350 & 0.6097 & 0.1378 & 0.6207 & 0.1401 & 0.6311 & 0.1438 \\
        DropoutNet& \textbf{0.5989} & \textbf{0.1352} & 0.6098 & 0.1374 & 0.6203 & 0.1396 & 0.6302 & 0.1435 \\
        Meta-E& 0.5982 & 0.1346 & 0.6093 & 0.1377 & 0.6195 & 0.1400 & 0.6294 & 0.1428 \\
        MWUF& 0.5986 & 0.1348 & 0.6082 & 0.1374 & 0.6184 & 0.1399 & 0.6279 & 0.1429 \\ \cdashline{1-9}[1.5pt/2pt]
        \name(Init Only)& 0.5987 & 0.1350 & 0.6098 & 0.1376 & 0.6204 & 0.1398 & 0.6306 & 0.1432 \\ 
        \name & 0.5978 & 0.1347 & \textbf{0.6198} & \textbf{0.1408} & \textbf{0.6308} & \textbf{0.1477} & \textbf{0.6380} & \textbf{0.1503}  \\ \hline
        \multicolumn{9}{l}{\textit{\textbf{Dataset: Taobao Ad \& Backbone: Wide\&Deep}}} \\
        Wide\&Deep& 0.6081 & 0.1360 & 0.6129 & 0.1427 & 0.6207 & 0.1455 & 0.6287 & 0.1484 \\
        DropoutNet& \textbf{0.6095} & 0.1359 & 0.6184 & 0.1427 & 0.6246 & 0.1454 & 0.6312 & 0.1474 \\
        Meta-E& 0.6082 & 0.1378 & 0.6122 & 0.1443 & 0.6190 & 0.1477 & 0.6259 & 0.1506 \\
        MWUF& 0.6089 & \textbf{0.1382} & 0.6125 & 0.1423 & 0.6210 & 0.1457 & 0.6285 & 0.1483  \\ \cdashline{1-9}[1.5pt/2pt]
        \name(Init Only)& 0.6027 & 0.1359 & 0.6065 & 0.1429 & 0.6163 & 0.1471 & 0.6232 & 0.1496 \\
        \name & 0.6051 & 0.1368 & \textbf{0.6220} & \textbf{0.1457} & \textbf{0.6290} & \textbf{0.1495} & \textbf{0.6336} & \textbf{0.1511}  \\ 
        \whline
\end{tabular}
\vspace{-0.3cm}
\end{table}

\begin{figure*}
    \centering
    \scalebox{0.9}{
    \includegraphics[width=\linewidth]{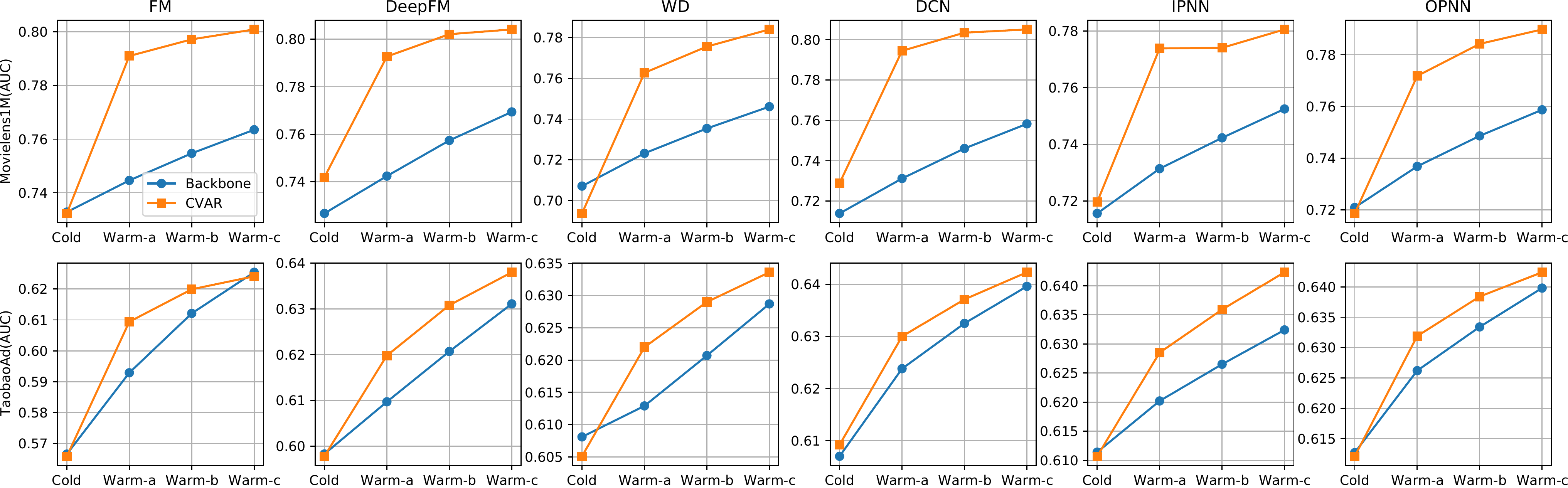}}
    \caption{AUC curves through warming-up on two datasets, over six popular backbone models, three runs for each.}
    \label{fig:compatibility}
    \vspace{-0.4cm}
\end{figure*}

\subsection{Offline Experiments}
\subsubsection{Experiment Setup}
In this section, we will introduce the offline experiment setup.

\textbf{Public Datasets.} For offline experiments, We evaluate \name on the following two public datasets MovieLens-1M\footnote{http://www.grouplens.org/datasets/movielens} and Taobao Ad\footnote{https://tianchi.aliyun.com/dataset/dataDetail?dataId=56}. 
\begin{itemize}
\item MovieLens-1M: One of the most well-known recommendation benchmark dataset. The data consists of 1 million movie ranking instances over thousands of movies and users. Features of a movie include its title, year of release, and genres which are seen as item side information. Titles and genres are lists of tokens.  Each user has features including the user’s ID, age, gender and occupation. 
\item Taobao Display Ad Click: It randomly samples 1140000 users from 26 million ad display / click records on Taobao~\footnote{https://www.taobao.com} website to construct the dataset. 
We take each Ad as a item for CTR prediction, with 4 categorical attributes as side information, including category ID, campaign ID, brand ID, advertiser ID. Each user has features including Micro group ID, cms\_group\_id, gender, age, consumption grade, shopping depth, occupation and city level.
\end{itemize}

\textbf{Dataset Split.} To demonstrate the recommendation performance in both cold-start and warm-up phases, we conduct the experiments by splitting the datasets following \cite{pan2019warm} and \cite{zhu2021learning}. We divide items into two groups, old and new based on their frequency, where items with more than $N$ labeled instances are old and others are new. We use $N$ of 200 and 2000 for Movielens-1M and Taobao Ad data. Note that the ratio of new items to old items is approximately 8:2, which is similar to the definition of long-tail items~\cite{chen2020esam}. Besides, new item instances sorted by timestamp are divided into four groups denoted as warm-a, -b, -c, and test set following \cite{pan2019warm} and \cite{zhu2021learning}.

\textbf{Backbones and Baselines.} Because \name is model-agnostic, it can be applied to various existing models in the Embedding \& MLP paradigm. Thus we conduct experiments upon the following representative backbones: FM~\cite{rendle2010factorization}, DeepFM~\cite{guo2017deepfm}, Wide\&Deep~\cite{cheng2016wide}, DCN~\cite{wang2017deep}, IPNN~\cite{qu2016product}, OPNN~\cite{qu2016product}. Meanwhile we choose some \textbf{S}tate-\textbf{O}f-\textbf{T}he-\textbf{A}rt(SOTA) methods for the item cold-start problem as baselines: DropoutNet~\cite{volkovs2017dropoutnet}, Meta embedding(Meta-E)~\cite{pan2019warm}, MWUF~\cite{zhu2021learning}. We reproduce each baseline based on open source code or their publications if the code is unavailable. We open all of the related source code on Github~\footnote{https://github.com/BestActionNow/CVAR}.

\textbf{Implementation Details.} For a fair comparison, we use the same setting for all methods. The MLPs in backbones and cold-start modules use the same structure with two dense layers (hidden units 16). The embedding size of each feature is fixed to 16. Learning rate and mini-batch size are set to 0.001 and 2048 respectively.
At the inference stage, $x_{freq}$ of new emerging items is set to the largest item frequency in the corresponding dataset to generate warmed-up embeddings.
Training is done with Adam~\cite{kingma2014adam} optimizer over shuffled samples. In experiments, we firstly use old item instances to pretrain the backbone model as well as the cold-start module and evaluate on the test set(Initialization phase). Then we in turn feed warm-a, -b, -c data  to train the backbone or \name and evaluate models on the test set step by step. We take the AUC score~\cite{ling2003auc} and the F1 score~\cite{huang2015maximum} as evaluation metrics.

\begin{table}[]
    \small
    \setlength{\abovecaptionskip}{3pt}
    \caption{\name performance(AUC) with different $x_{freq}$ on Movielens1M and Wide\&Deep, three runs for each. The best improvements are highlighted in bold.}
    \label{tbl-x-freq}
    \centering
    \renewcommand{\arraystretch}{1.2}
    \begin{tabular}{c|cccc}
        \whline
        $x_{freq}$ & Cold & Warm-a & Warm-b & Warm-c \\ \hline
        0.01 & 0.6936 & 0.7627 & 0.7756 & 0.7839 \\
        0.1 & 0.6939 & 0.7629 & 0.7756 & 0.7837 \\
        0.25 & 0.6946 & 0.7627 & 0.7754	 & 0.7842 \\
        0.5 & 0.6956 & 0.7638 & 0.7754 & 0.7844 \\
        1.0 & \textbf{0.6973} & \textbf{0.7649} & \textbf{0.7757} & \textbf{0.7845} \\
        \whline
\end{tabular}
\vspace{-0.5cm}
\end{table}

\subsubsection{Experimental Results}
We compare the cold-start effectiveness of \name with backbone and three SOTA cold-start baselines including DropoutNet, Meta-E, MWUF. Meanwhile, we choose two most famous CTR prediction methods as backbones, DeepFM and Wide\&Deep. 
To evaluate the quality of initial embedding generated by \name and further prove the effectiveness of \name in leveraging the changing id embedding for better warm-up, we conduct a version of contrast experiment for \name denoted as \name(Init Only), where \name only plays a role in initialization phase and is disabled in following three warm-up phases. We conduct experiments on two datasets and evaluate the mean results over three runs.

The main experimental results are shown in Table~\ref{tbl-cold-start}. Notice that \name(Init Only) outperforms other baselines in most cases, which indicates a high quality of initial embeddings generated by \name. Moreover, superior performance of \name(Init Only) proves that distribution alignment in latent space is better than directly mapping side information to item id which is adopted in Meta-E and MWUF.
Except for a better initialization, \name can significantly improve the prediction performance in warm-up stages which is proven by the outstanding results of \name in Table~\ref{tbl-cold-start}. This phenomenon demonstrates that \name can indeed produce high quality warmed-up embedding based on the evolving item ID embedding learned by backbone in warm-up stage.

\textbf{Method compatibility.}
Since \name is model-agnostic, we conduct experiments in more scenarios to verify its compatibility. Results on six popular backbones and two datasets in Figure~\ref{fig:compatibility} demonstrate the compatibility and robustness of \name.

\textbf{Comparison of different frequency condition}
 As mentioned before, $x_{freq}$ is set as the condition of decoder in \name and has a direct impact on the distribution of ID embedding. Considering the final $x_{freq}$ is uncertain for new emerging items at inference stage, we compare the performance of \name with different $x_{freq}$. 
 Because  $x_{freq}$ is normalized before using, we conduct five experiments with $x_{freq}$ equal 0.01, 0.1, 0.25, 0.5 and 1. Experimental results are shown in Table~\ref{tbl-x-freq}. It's shown that performance of \name increases gradually with the increase of $x_{freq}$, which explains why we set $x_{freq}$ as a huge value at inference stage for cold start.

\subsection{Online A/B tests}
To verify the effectiveness of \name, we further conduct online A/B tests~\cite{kohavi2017online} for 7 days on Tencent News recommendation platform\footnote{https://news.qq.com/}. As usually adopted in industry recommendation, the whole coarse-to-fine recommendation progress can be divided into four stages : candidate generation, coarse-grained ranking, fine-grained ranking, and re-ranking.
In our industrial scenario, \name is applied to the ranking stage whose backbone is MMOE~\cite{ma2018modeling,tang2020progressive}. Consistent with the four phases in offline experiments, we group online items into four groups: cold, warm-a, -b, -c, with increasing exposed frequency. In our system, there are two mediums of news: Article and Video. We focus on the following metrics~\cite{gunawardana2009survey} (from high importance to low): Exposure Rate, Watch Time, Page(Video) Views. Exposure Rate measures the distribution percentage of the item group. Watch Time and Page(Video) Views reflect how users are attracted by recommended content.

We show the online results on various item groups in Table~\ref{tbl:online}. It's apparent that metrics of cold items are significantly improved, which proves the effectiveness of \name. As expected, the gain of \name gradually diminishes as the exposed frequency increases. Moreover, exposed items' Gini coefficient~\cite{zhou2010impact} on various item categories reduces from 0.7413 to 0.7369 after applying \name, which indicates \name could alleviate the Matthew Effect~\cite{wang2018quantitative} to some extent.

\begin{table}[]
\caption{Online A/B results of item groups with increasing warm-up level: \textit{cold}, \textit{warm-a}, \textit{warm-b} and \textit{warm-c}. Red results mean they are statistically significant (whose p-value in hypothesis testing~\cite{biau2010p} is stable less than 0.05.)}
\label{tbl:online}
\centering
\renewcommand{\arraystretch}{1.2}
\small
\setlength{\abovecaptionskip}{-1cm}
\scalebox{0.9}{
\begin{tabular}{c|ccccc}
\toprule[2pt]
Metrics            & Cold & Warm-a & Warm-b & Warm-c & Total     \\ \hline
Exposure Rate      & \textcolor{red}{+1.48\%} & -0.21\% & -0.04\% & +0.17\% & -   \\ \hline
Watch Time         & \textcolor{red}{+2.49\%} & \textcolor{red}{+2.90\%} & +1.40\% & +0.39\% & +0.38\%    \\
Article Watch Time & \textcolor{red}{+2.39\%} & \textcolor{red}{+4.51\%} & +2.08\% & +0.16\% & +0.13\%   \\
Video Watch Time   & \textcolor{red}{+2.60\%} & +1.78\% & +0.72\% & +0.68\% & +0.66\%    \\ \hline
Total Page Views         & \textcolor{red}{+4.46\%} & +2.87\% & +1.42\%& +0.62\% & +1.09\%    \\
Article Page Views & \textcolor{red}{+3.58\%} & \textcolor{red}{+3.37\%} & +1.74\%& +0.31\% & +0.82\%     \\
Video Views   & \textcolor{red}{+5.84\%} & +2.25\% & +1.05\%& +1.01\% & +1.35\%     \\
\bottomrule[2pt]
\end{tabular}}
\vspace{-0.6cm}
\end{table}

\section{Conclusion}

In this paper, we proposed the model-agnostic \name for item cold-start
which uses latent variables to learn a distribution over side information and generates desirable ID embeddings using a conditional decoder.
For better cold-start performance, \name not only learns the pattern in historical data but also
leverages the continuously updating item ID embedding from the backbone to generate enhanced warmed-up embeddings with superior quality. From the aspect of industrial practice, we claim that additional strict data requirements of cold-start methods will make the deployment process rather difficult in online scenario. Thus \name is designed to be trained with the same raw samples as training main prediction model. Note that the proposed \name is a general framework that can be applied to various backbones. Finally, 
extensive offline experiments on public datasets and online A/B tests show the effectiveness and compatibility of \name.

\clearpage
\bibliographystyle{ACM-Reference-Format}
\bibliography{custom}

\end{document}